# Existence of nearest-neighbor and variable range hopping in $Pr_2ZnMnO_6$ oxygen-intercalated pseudocapacitor electrode

Moumin Rudra[#,1], H.S.Tripathi[1], Alo Dutta[2] & T. P. Sinha[1]

[1]Department of Physics, Bose Institute, 93/1, APC Road, Kolkata, India – 700009.
[2]Dept. of Condensed Matter Physics & Material science, S. N. Bose National Centre for Basic Sciences, Block-JD, Sector-III, Salt Lake, Kolkata – 700098
[#]Email id: *iammoumin@gmail.com*

## ABSTRACT

The X-ray diffraction pattern and Raman spectrum authenticate the monoclinic $P2_1/n$ structure of polycrystalline $Pr_2ZnMnO_6$ (PZM) synthesized by the solid-state reaction technique. The X-ray photoemission spectrum suggests the existence of oxygen vacancy in PZM, which plays a crucial role in electrical conduction as well as in electrochemical behavior. The existence of two different conduction mechanisms (such as nearest-neighbor hopping (NNH) and Mott's variable range hopping (MVRH)) is observed from the investigation of dc conductivity studies, dielectric relaxation, and impedance analysis of PZM. It is found that below 580 K, there is a transition from NNH to MVRH due to the decrease of activation energy. The constant phase element model is used to correlate the Nyquist plot and electric modulus of PZM. The bulk PZM electrode exhibits intercalated pseudocapacitive nature which offers an efficient specific capacitance of 69.14 F/g at a charge and discharge currents of 2 A/g in an aqueous solution of 1 M KOH. The charge storage mechanism of the PZM electrode has been analyzed and discussed.

## I. INTRODUCTION

The supercapacitor, a high-capacity capacitor, which works as a link between traditional capacitors and rechargeable batteries, with long cycle life, fast charge/discharge rate, and high-power density has optimistic scenarios [1-6]. But its low energy density has limited its real-life application. To remove this kind of shortfall, a lot of work has been committed to designing the electrode materials for high energy density supercapacitor [7-9]. As stated by the charge storage mechanism, these electrode materials can be specified into two groups: double-layer capacitor [DLC] [10] and pseudo-capacitors [11]. DLC materials can store charges without any electrochemical reaction, though charge accumulation at the electrode-electrolyte interface. On the other side, pseudo-capacitor materials can store charges through the redox reactions or ion intercalation at the edge of the electrolyte and the surface of electrode. These pseudo-capacitor materials exhibit both high energy density and high-power density. These electrode materials are four types: Conducting polymers, Carbon-based materials, Hydroxides [12], and transition metal oxides. Conducting polymers exhibit poor electrochemical stability [13], whereas Carbon-based materials have good thermal and chemical stability, but it exhibits low energy density [14]. On the other side, hydroxides and transition metal oxides have better electrochemical stability and higher energy density [1].

The transition metal-based perovskite oxides are now being examined as potential candidates for supercapacitor applications due to their thermal and chemical stability and ability to contain oxygen vacancies. In the year 2014, Mefford *et al.* [15] proposed the charge storage mechanisms in $LaMnO_3$ based on oxygen anion-intercalated pseudocapacitor. Recently, Liu *et al.* [16] also found an excellent oxygen-anion based intercalated pseudocapacitive behavior in Pr-based double perovskite $PrBaMn_2O_6$. This study also suggests that the doping of $Ba^{2+}$in place of $Pr^{3+}$in $PrMnO_3$ creates oxygen vacancies in the material, which is the main requirement to achieve a high-performance supercapacitor. Here, we have replaced half of the $Mn^{3+}$ by $Zn^{2+}$ in $PrMnO_3$, aimed to oxidize $Mn^{3+}$ into $Mn^{4+}$ due to charge neutrality in the system. By consulting different works of literature [17, 18], it is evident that the coexistence of $Mn^{3+}$ with the $Mn^{4+}$ in different Mn-based perovskite oxides. The existence of $Mn^{3+}$ in the system will bring oxygen vacancies in the material to maintain charge neutrality, which will play a crucial role in electrochemical properties. The charge storage mechanisms in $Pr_2ZnMnO_6$ (PZM) electrodes should be scrutinized to develop the innovative pseudocapacitor electrodes with robust cyclic stability and a high energy density. For the first time, the material PZM synthesized successfully. In this article, we have employed X-ray photoelectron spectroscopy to finding the oxygen vacancy in the PZM. After that, the electrical conductivity, electrical impedance and electrochemical impedance spectroscopy were used to investigate its electrical and electrochemical transport properties of PZM, respectively. Also, we will discuss how effectively this material is used for two different types of conduction. This material PZM opens a new area of research in the field of electrochemical application for the upcoming years.

## II. EXPERIMENTAL

PZM in powder form was synthesized by a standard solid-state reaction technique at 1573 K in the air for 14 hrs. using $Pr_2O_3$ (Sigma-Aldrich, 99.9%), ZnO (Merck, 99%), and $MnO_2$ (Loba Chemie, 99.99%) as the starting materials. These starting materials were mixed in presence of acetone (Merck, 99%) for 8 hr periodically. The crystallographic structure of calcined powder of PZM was studied using powder X-ray diffractometer (Rigaku Ultima IV) [40 kV, 40 mA, Cu-$K_\alpha$ ($\lambda$ = 1.5406 Å) radiation] in the 2θ range of 10° – 120° by the scanning rate of 0.02° per step at room temperature. A dual-position graphite diffracted beam monochromator has been used with a scintillation counter detector. The collected XRD data were refined with the help of the FULLPROF program [19] using the Rietveld method [20]. During the Rietveld refinement, the pseudo-Voigt functions were used to fit peak shapes and the background was fitted using a 6-coefficient polynomial function. After confirmation of material structure, we used the polyvinyl alcohol (PVA) (5%) with the calcined powders and the mixture was pressed inside the cylindrical pelletizer. The pressed pellets were sintered at 1623 K to obtain dense PZM pellets with thickness, $d \approx 1$ mm & diameter $\approx 8$ mm. The microstructural image of the sintered pellet was captured using a scanning electron microscope (SEM) (FEI Quanta 200). The unpolarized Raman spectrum of the PZM was measured by a LABRAM HR 800 system at room temperature. The sample was excited using a laser of wavelength 488 nm (Ar-ion laser) with a 100x objective (NA 0.9). The X-ray photoelectron spectroscopy (XPS) analysis was executed on the MXPS system (Scienta Omicron, Germany) equipped with the EA 125 U5 hemispherical analyzer with an XM1000 monochromator and a CN10 charge neutralizer. The monochromatized Al-$K_\alpha$ radiation (1486.7 eV) with the output power set at 300W was used for the XPS analysis. To calibrate the binding energies (BE) of various elements in the sample, the C(1*s*) signal (284.9 eV) was hired as the reference. The impedance spectroscopy study of the sintered pellet was measured in the temperature range from 300 K to 700 K (frequency range ≈ 45 Hz to 5 MHz), using an LCR meter (HIOKI – 3532, Japan). Before starting the experiment, both bases of the measuring pellet were cleaned and the contacts were prepared from the thin silver paste. The real ($\varepsilon'$) and imaginary ($\varepsilon''$) parts of the complex dielectric constant $\varepsilon^*$ ($= \varepsilon' + j\varepsilon''$, where $\varepsilon' = C_s/C_0$ and $\varepsilon'' = \mathbb{G}/\omega C_0$) were obtained from the capacitance ($C_s$) and the conductance ($\mathbb{G}$), whereas the real (Z') and imaginary (Z'') parts of the complex impedance Z* ($= Z' - jZ''$, where $Z' = Z\cos\varphi$ and $Z'' = Z\sin\varphi$) were obtained from the impedance (Z) and the phase angle ($\varphi$) [where ω is the angular frequency ($\omega = 2\pi\nu$), ν is the measured frequency and $j = \sqrt{(-1)}$. $C_0 = \varepsilon_0 A/d$ is the parallel plate capacitance of empty cell; A is the base area and *d* is the thickness of pellet]. The ac electrical conductivity $\sigma$ ($= \mathbb{G}d/A$) was measured from the $\mathbb{G}$. The dc conductivity values are extracted from I-V characteristic with the help of Keithley (USA) Data Acquisition Multimeter System.The room temperature electrochemical properties of the PZM electrodes were studied using the cyclic voltammetry (CV), the galvanostatic charge-discharge (GCD) and the electrochemical impedance spectroscopy (EIS) by employing PGLyte conventional three-electrode electrochemical cell consisting of PZM over Nickel foam as working electrode, highly pure platinum (Pt) mesh (1cm x 1cm) as the counter electrode and the saturated Ag/AgCl as the reference electrode in a 1 M KOH electrolyte solution. The active working electrode material was prepared by mixing graphite powder (10 wt%), polyvinylidene fluoride (PVDF, 5 wt%), and PZM powders (85 wt%) in 200μL N-Methyl-2-pyrrolidone (NMP) under the sonication for 30 min to obtain the gel. The gel was then dropped cast on the one flat surface of the Nickel foam and dried in vacuum at 343 K for 8 hrs. The average mass loading density of PZM on the Nickel foam was about 1 mg/cm$^2$. The CV measurements were carried out at various scan rates of 2, 5, 10, 20, 50 and 100 mV/s, within a potential range -0.8 to 0.5 V. The specific capacitance ($C_{sp}$) was calculated using GCD experiments at various current densities of 2, 3, 4, 5, 6, 8, 10 A/g. The alternating current (AC) electrochemical impedance was studied in the frequency range of 100 mHz to 100 kHz at room temperature.

## III. RESULTS AND DISCUSSION

### A. Structural analysis

The powder XRD profile for PZM at room temperature is presented in FIG. 1 (a). The red open circles represent the observed data, whereas the black solid line is the calculated refined data for PZM. The difference (blue solid line) between the observed data and calculated results is small, signifying the eminence of refinement with $\chi^2$= 1.01. The reliability factors are found to be $R_{exp}$ = 6.95, $R_p$ = 5.46, and $R_{wp}$ = 6.97. The refined parameters are listed in TABLE 1. The existence of the super-lattice diffraction peak (101) at 2θ ~ 20° and (022), (202), ($\bar{2}02$) peaks around 40° in the XRD profile indicates the monoclinic $P2_1/n$ crystal structure [21]. Since the monoclinic $P2_1/n$ structure is a centrosymmetric subgroup of the aristotype cubic $Fm\bar{3}m$, the transformation matrixes for the cubic to monoclinic phases are given in TABLE 1. The mode decomposition of the cubic $Fm\overline{3}m$ to the monoclinic $P2_1/n$ symmetry breaking for the material PZM and the irreducible representation modes are shown in the Supplementary TABLE ST1. The fitting result shows monoclinic $P2_1/n$ phase with $a$ = 5.4578 Å, $b$ = 5.5295 Å, $c$ = 7.7255 Å and $\beta$ = 90.089º. The value of $\beta$ (≠ 90º) which is slightly different from the other angles $\alpha$ and $\gamma$ (= 90º) indicates the metric structure to be pseudo-orthorhombic. A schematic representation of the PZM unit cell is shown in FIG. 1 (b). The distribution of the atoms at the crystallographic positions *4e* (xyz) for $Pr^{3+}$, *2c* (00½) for $Zn^{2+}$, *2d* (½0½) for $Mn^{4+}$ and *4e* (xyz) for $O^{2-}$. Each $Zn^{2+}$ and $Mn^{4+}$ atoms bonded by the six $O^{2-}$ atoms institute the $ZnO_6$ and $MnO_6$ octahedra, respectively. There is no mixing of the sites between two B-site cations Zn and Mn during the refinement process, which is confirmed by interchanging Zn at Mn-site and vice versa. In this monoclinic structure, the smaller A-site cation (i.e., Pr atom) forces the $ZnO_6$ and $MnO_6$ octahedron to tilt to optimize Pr–O bond distances. The B-site octahedron $ZnO_6$ and $MnO_6$ are fully ordered and alternate along with the three-dimensional network of the material PZM. Each $ZnO_6$ octahedron is linked with six $MnO_6$ octahedra and vice versa. The $ZnO_6$ and $MnO_6$ octahedra having 1:1 alternate ordering in monoclinic $P2_1/n$ structure are tilted in-phase along with the (101) direction of the pseudocubic cell and also in anti-phase along with the (010) and (001) directions, which corresponds to the Glazer's notation $a^-a^-c^+$. The transition metal (Zn/Mn) – oxygen bond lengths, Zn–O–Mn bond angles, and bond-valence sum (BVS) values for PZM are also shown in TABLE 1. We have calculated the distortion index ($\Delta_d$) using the equation below:

$$\Delta_d = \frac{1}{6}\sum_{n=1}^{6}\left[\frac{d_n - \langle d\rangle}{\langle d\rangle}\right]^2 \qquad (1)$$

where $d_n$ is an individual B–O bond length and $\langle d\rangle$ is the average B–O bond length. The value of $\Delta_d$ greater than $10^{-3}$ indicates the significant distorted bond length in the $BO_6$ octahedra. For PZM, the value of $\Delta_d$ is found to be $6.5\times10^{-4}$, which suggests nearly rigid octahedral tilting. The tilt angle is calculated from the Zn–O–Mn bond angle (δ) as [(180º–

δ)/2] and it is found to be 11.67°. The global instability index ($G$) is a crucial parameter to check the stability of the monoclinic structure [22] and it is calculated using the following expression,

$$G = \left[\frac{\sum(\mathrm{BVS} - V_A)^2}{N}\right]^{1/2} \tag{2}$$

where $V_A$ is the valency of an individual atom, $N$ is the total number of atoms present in the unit cell. The value of $G$ is found to be 0.095, which suggests a stable monoclinic structure of PZM. The tolerance factor is also calculated using the equation, $t = (r_{Pr} + r_O)/\sqrt{2}(r_B + r_O)$, here $r_{Pr}$, $r_O$, and $r_B$ are the ionic radii of praseodymium (Pr), oxygen (O) and B-site cation, respectively. The $t$ value is found to be 0.896, which also supports the lower symmetric monoclinic structure of PZM. The SEM image of the PZM pellet, shown in FIG. 1 (c), indicates the high density of the material as well as grains (Gs) of various sizes and shapes. The size of the Gs differs from 1 μm to 2μm. The observed density is calculated utilizing the Archimedes' principle and its value is found to be 6.74 g/cc ($\times 10^3$ kg/m$^3$), which is 95% of the theoretical value 7.09 g/cc ($\times 10^3$ kg/m$^3$).

## B. Raman spectroscopy

The room temperature unpolarized Raman spectrum of PZM is shown in FIG. 2 (a). It is observed that the peaks at 94, 154, 235, 281, 429, 541, 672, and 858 cm$^{-1}$ ($\times 10^2$ m$^{-1}$) are the dominating bands with the other weakly bands. In the region 50 – 200 cm$^{-1}$ ($\times 10^2$ m$^{-1}$), a large number of bands are indicating the lower symmetric structure of PZM. As discussed in Sec. III (A), PZM crystallizes in a monoclinic $P2_1/n$ ($C^5_{2h}$) structure, which contains two molecules per unit cell (Z = 2). The monoclinic $P2_1/n$ (Glazer's notation $a^-a^-c^+$) structure is derived from the cubic aristotype $Fm\bar{3}m$ ($a^0a^0a^0$) symmetry by anti-phase and in-phase tilts of $BO_6$ octahedra. The aristotype $Fm\bar{3}m$ symmetry represents the 1:1 ordered lattice arrangements in DPOs. This arrangement allows a classification of the normal modes at the Brillouin zone center as:

$$\Gamma = A_g \oplus E_g \oplus F_{1g} \oplus 2F_{2g} \oplus 5F_{1u} \oplus F_{2u} \quad (3)$$

Out of 11 phonon modes, there are four ($A_g$, $E_g$, $2F_{2g}$) Raman active, four ($4F_{1u}$) infrared (IR) active, one ($F_{1u}$) acoustic, and two ($F_{1g}$, $F_{2u}$) silent modes. In our material, Zn and Mn ions should occupy the *2c* and *2d* Wyckoff sites of $C_i$ symmetry, and the A cation (Pr) and anion (O ions) would be in *4e* sites of $C_1$ symmetry. This factor group analysis leads to the distribution of irreducible representations of the $C_{2h}$ point group as below:

$$\Gamma = 12A_g \oplus 12B_g \oplus 18A_u \oplus 18B_u \quad (4)$$

where 24 ($12A_g \oplus 12B_g$) Raman active modes are expected in the monoclinic material [23]. The correlation between both phases and their prototype phase is shown in the Supplementary TABLE ST2. With the help of factor group analysis, we determine the distribution of zone-center vibrational modes in terms of the respective representations of the $C_{2h}$ point group, and the results are tabulated in the Supplementary TABLE ST3. The spectrum containing several phonon modes of the monoclinic structure of PZM is fitted applying the Lorentzian profile as shown in FIG. 2 (b) – 2 (d). The observed phonon modes with frequency and the corresponding FWHM parameters after deconvolution of the Raman spectrum are tabulated in TABLE 2. Considering only 24 Raman modes, the $P2_1/n$ irreducible representation can be written as

$$\Gamma^{Raman} = \nu_1(A_g \oplus B_g) \oplus \nu_2(2A_g \oplus 2B_g) \oplus \nu_5(3A_g \oplus 3B_g) \oplus L(3A_g \oplus 3B_g) \oplus T(3A_g \oplus 3B_g) \quad (5)$$

where *L* and *T* represent the librational and the translational modes of the Pr-atom and $\nu_1$&$\nu_2$ correspond to the symmetric & antisymmetric stretching of $BO_6$ octahedra, respectively. The symmetric bending mode of $BO_6$ octahedra denotes by $\nu_5$. Since the monoclinic $P2_1/n$ structure is a centrosymmetric subgroup of the aristotype cubic $Fm\overline{3}m$ structure, just Raman active modes of $BO_6$ octahedra are considered. In this case, the corresponding phonon modes are determined by $B^{II}$-O and $B^{IV}$-O binding forces. Here $B^{IV}$-O bond is the strongest bond and also $B^{IV}$ is lighter than $B^{II}$ atom. So, the higher frequencies should be assigned to $\nu_1$, $\nu_2$ and $\nu_5$ vibrations of the $B^{IV}$-O octahedron with $\nu_1>\nu_2>\nu_5$. Taking into account, the breathing vibration $\nu_1$ ($A_g$ mode) at 858 $cm^{-1}$ ($\times 10^2$ $m^{-1}$) of PZM, is allotted to the symmetric stretching of $MnO_6$ octahedra. Another $\nu_1$ peak at 797 $cm^{-1}$ ($\times 10^2$ $m^{-1}$) is also symmetric stretching of $MnO_6$ octahedra assigned to $B_g$ mode. The antisymmetric stretching of the $MnO_6$ octahedra allotted to $A_g$ and $B_g$ ($\nu_2$ peaks) modes which appear at 541, 608, and 672 $cm^{-1}$ ($\times 10^2$ $m^{-1}$) for the monoclinic structure. The internal bending motions of the $MnO_6$ octahedra, $\nu_5$ peaks fall in the range of 380 – 500 $cm^{-1}$ ($\times 10^2$ $m^{-1}$). Proceeding to the lower wavenumbers, the *L* modes are the expected peaks to appear. The peaks in the range of 160 – 350 $cm^{-1}$ ($\times 10^2$ $m^{-1}$) are too weak and assigned as the *L* modes of the Pr – cation. Most of the perovskite compounds [24] reported that the peaks in the range of 350 – 420 $cm^{-1}$ ($\times 10^2$ $m^{-1}$) cannot be distinguished as either *L* modes or $\nu_5$ modes of the material. In $P2_1/n$ symmetry, six translational (*T*) modes are expected, via a triplet splitting of the $A_g$ and $B_g$ modes. The *T* modes of the Pr-cation at low wavenumbers are found below 160 $cm^{-1}$ ($\times 10^2$ $m^{-1}$). A total of 19 out of 24 phonon modes have been found in the Raman spectrum of PZM because of the small correlation field splitting of the polycrystalline PZM. The representative displacement vectors of the $MnO_6$ are shown in the Supplementary FIG. SF1.

### C. X-ray photoemission spectroscopy

To confirm the mixed-valence state of the cations and the oxygen vacancies in PZM, the XPS spectra of Pr (3*d*), Mn (2*p*), and O (1*s*) are displayed in FIG. 3 (a) – 3 (c), respectively. In FIG. 3 (a), the observed Pr*3d* XPS result is similar to that of $Pr_2O_3$ as reported by Burroughs *et al.* [25]. The observed *3d* XPS spectrum shows two main peaks at 932.8 eV [1 eV = 1.602 × $10^{-19}$ J] and 953.2 eV, which are assigned to be Pr ($3d_{5/2}$) and Pr ($3d_{3/2}$) levels, respectively. The difference between the two main peaks is about 20.4 eV, which is commonly reported for the spin-orbit splitting of the $3d_{5/2}$ and $3d_{3/2}$ levels [26]. According to Moulder *et al.* [27], these peak positions are expected for the $Pr_2O_3$ and these are quite different from the $PrO_2$ peak positions. The $3d_{5/2}$ and $3d_{3/2}$ levels show two satellite peaks at low binding energies (928.2 eV and 948.6 eV, respectively) respect with two main peaks, which may be attributed to the well-screened $4f^3$ final state [28]. Another cusp at 956.9 eV (represented by the asterisks in FIG. 3 (a)) represents an extra structure at $3d_{3/2}$ due to the multiplet effect, which is absent in $3d_{5/2}$ [29]. These results suggested that the Pr ions in the synthesized materials are in the $Pr^{3+}$ state. The XPS spectra of Mn (2p) is shown in FIG. 3 (b). Both the Mn $2p_{3/2}$ and $2p_{1/2}$peaks can be deconvoluted into two peaks by Lorentzian curve fitting, which suggests the existence of two different Mn states in PZM. The values of the binding energy at 641.6 eV and 653.8 eV suggest the existence of $Mn^{4+}$ state for $2p_{3/2}$ and $2p_{1/2}$ regions, respectively. On the other hand, another two peaks at 641.2 eV and 653.4 eV indicate the existence of $Mn^{3+}$ state. For the $2p_{3/2}$ peak, we have found that the percentage of $Mn^{3+}$ existence in PZM is about 13% from the peak area. This low amount of $Mn^{3+}$ state leads either $Pr^{3+}$ undergoes into $Pr^{4+}$ or creates oxygen vacancies, which may lead the conduction dynamics in the material. Since the XPS spectrum of Pr suggests the $Pr^{3+}$ state in PZM, the existence of $Mn^{3+}$ confirms the oxygen vacancies in the material. FIG. 3 (c) shows the XPS spectra of O-*1s*, which is deconvoluted into two Lorentzian peaks, which suggests the existence of two types of oxygen-related species. The O-*1s* spectra show similar results as reported in $Pr_2O_3$ [26]. The peak at 529.1 eV is assigned as the peak due to the structural oxygen, whereas the peak at 531.3 eV may be due to the existence of chemisorbed oxygen-related species, such as the hydroxyl group ($OH^-$) or other species (CO, $CO_2$) at the surface of the sample [30-32] or the existence of oxygen vacancies [33, 34] in the material. In general, the existence of $Mn^{3+}$ ion along with $Mn^{4+}$ ion in PZM suggests that the oxygen vacancies will be required to maintain charge neutrality according to the equation given below:

$$4Mn^{4+} + O^{2-} \rightarrow 4Mn^{4+} + 2e^{-}/\square + \tfrac{1}{2}O_2 \rightarrow 2Mn^{4+} + 2Mn^{3+} + \square + \tfrac{1}{2}O_2 \quad (6)$$

This $\square$ represents an empty position, which is creating from the removal of $O^{2-}$ ion in the material. It can be assumed that an oxygen vacancy created along with two $Mn^{3+}$ ions from the above Eq. 6. Therefore, the concentration of the oxygen vacancies in the PZM is calculated and found to be 4% [35]. Naldoni *et al.* [36] have claimed about 5% oxygen vacancies in their material due to the fast cooling rate during calcination, which is the main reason to retain oxygen vacancies in the material.

**D. Conductivity study**

The variation of the ac conductivity [$\log\sigma_{ac}$ vs $\log\omega$] with frequency at various temperatures is shown in FIG. 4. We can split the ac conductivity ($\sigma_{ac}$) spectra into two distinct regimes. The first regime (i.e., regime I), where the $\sigma_{ac}$ is independent of frequency, suggesting the dc conductivity ($\sigma_{dc}$) type behavior. The second one (i.e., regime II), where the $\sigma_{ac}$ strongly depends on frequency as both increases simultaneously. These two regimes can ~~be~~ explain~~ed~~ using the Jump relaxation model, which suggests the ionic conduction process in solids based on successful and unsuccessful hopping of the charge carriers [37, 38]. The successful hopping of charge carriers can described as the charge carrier jumps at the nearest-neighboring site~~s~~ from its initial site (i.e., forward jump) and the charge carriers elect to stay in the nearest-neighboring (new) site. On the other side, the unsuccessful hopping of charge carriers can described as the charge carrier jumps at the nearest-neighboring site~~s~~ and then return back to its initial site (i.e., reverse jump). In this respect, the ac conductivity spectra at regime I is associated with the successful hopping of charge carriers, whereas the unsuccessful hopping dominates over the successful hopping at regime II.

The dc conductivity ($\sigma_{dc}$) as a function of temperature for PZM is shown in FIG. 5 (a). Inset shows $\log \sigma_{dc}$ versus temperature plot. The plot shows a sharp change in the conductivity slop, which suggests the existence of dissimilar hopping models in the measured temperature range. To report the type of conduction mechanisms in perovskite oxides at various temperatures, various hopping models are projected. The nearest-neighbor hopping (NNH) is one of the known conduction mechanisms in various perovskite oxides [39-41]. In this conduction mechanism, the activation energy has a constant single value, which leads the charge carriers to hop on the nearest-neighbor site from the initial site. To identify the hopping models involved in the conductivity mechanism, we have plotted the observed $\sigma_{dc}$ data using the NNH model as shown in FIG. 5 (b) and the equation used is written as

$$\sigma_{dc}T = \sigma_{\alpha} e^{-\left({E_a}/{k_B T}\right)} \quad (7)$$

Here $\sigma_{\alpha}$ is the pre-exponential factor, $E_a$ is the activation energy, $k_B$ is the Boltzmann constant, and $T$ is the absolute temperature. The curve is fitted well at the high-temperature regime (above 580 K), where it follows the linear NNH model. We have found the value of $E_a$ is 0.15 eV. But, below 580 K, the observed data show the non-linear behavior which suggests that there is more than one activation energy. This result points towards a different conduction dynamics presence in our material. In this condition, the charge carriers hops to a long distance energetically promising site with various $E_a$. In this region, the different $E_a$values are calculated using Eq. 8 and the variation of $E_a$ with temperature is shown in FIG. 5 (c).

$$E_a = -\frac{d[\ln(\sigma_{dc})]}{d\left[{1}/{k_B T}\right]} \quad (8)$$

This figure shows the values of $E_a$ decrease with the decreasing temperature, which is not a sign of NNH model. It is well known that ~~at~~ the conductivity data for several semiconductors [42] deviate from the linear nature (i.e., NNH model) and obey the temperature dependency in the form [43].

$$\sigma_{dc} = \sigma_0 e^{-\left[T_0/T\right]^{\gamma}} \tag{9}$$

where $\gamma$ is either ¼ or ⅓ or ½ and depends inadequately on the temperature as well as on the other parameters. Mott's variable range hopping (MVRH) [$\gamma$ = ¼] is one type of hopping mechanism [44], where mean hopping length decreases with an increase in temperature. In this aspect, we can assume that MVRH is present below 580 K, where single $E_a$ cannot be extracted. FIG. 5 (d) shows the fitting results of the observed data with the MVRH formalism using Eq. 10

$$\sigma_{dc} = \sigma_0 e^{-\left[T_0/T\right]^{1/4}} \tag{10}$$

where $\sigma_0$ is a constant parameter and $T_0$ is the characteristic temperature (Mott's), which can be defined as follows

$$T_0 = \frac{18\alpha^3}{k_B N(E_F)} \tag{11}$$

Where the $N(E_F)$ is the density of states at the Fermi energy and "$1/\alpha$" be the localization length. The value of $T_0$ is extracted from the linear fit in FIG. 5 (d) and is found to be 54171311 K. The hopping energy $E_h$ ($T$) for a certain temperature $T$ is calculated using the Eq. 12 [45]

$$E_h(T) = \frac{1}{4} k_B T^{\frac{3}{4}} T_0^{\frac{1}{4}}. \tag{12}$$

The variation of the $E_h$ ($T$) with the temperature is shown in the inset of FIG. 5 (d) and it is witnessed that the $E_h$ ($T$) increases from 0.133 eV at 300 K to 0.204 eV at 580 K. The value of localization length ($1/\alpha$) is taken as 0.98 Å, which is estimated for the Mn-based perovskite as reported in previous literature [46]. This low value of $1/\alpha$ can be clarified by considering the charge carriers hopping between the localized acceptor states at the grain boundaries (GBs) and it plays the leading role in the conduction mechanism at this temperature regimes. Exactly similar behavior is suggested by many works of literature of materials like $Pr_{0.8}Ca_{0.2}MnO_3$ [46, 47]. The value of hopping length $R_h(T)$ is calculated by $\frac{3}{8\alpha}\left(\frac{T_0}{T}\right)^{\frac{1}{4}}$ [48]. The value of $R_h(T)$ decreases from 7.576 Å at 300 K to 6.425 Å at 580 K, by considering the value of $1/\alpha$. It is also witnessed that the minimum value of $R_h(T)$ is of the order of the average distance ($d$) of the manganese ions ($d$ = 6.42 Å), which also implies that 580 K is the maximum temperature for the MVRH to arise [49]. From these results, we can conclude that the $\sigma_{dc}$ points towards a transition from NNH to MVRH around 580 K with a decreasing $E_a$.

**E. Impedance analysis**

The impedance spectroscopy is a powerful technique in material science due to its capability to recognize the transport behavior in Gs and GBs. The Gs and GBs are the two main components of the microstructures of material and the correlations between Gs and GBs are also important to understand the complete electrical properties in the PZM. The Nyquist plots for PZM are shown in FIG. 6 (a) & 6 (b). These plots show two semicircular arcs having their center below the X-axis. The intersection on the X-axis at the low-frequency side gives the total resistance. Each curve passing through the origin (0, 0) at high frequency side, which suggesting Gs contribution in PZM. At a particular temperature, the plot shows two semicircular arcs suggesting that two different relaxation processes are involved in our measured temperature range. Out of these two different processes. one is for GBs contribution and another is for Gs contribution. The GBs are more resistive than Gs. Therefore, the effects of GBs are found on the low-frequency side, where the Gs contribution is visible on the high-frequency side.

To find the microstructural effect on electrical transport properties of PZM, we have employed a similar circuit model $[(R_G Q_G)(R_{GB} Q_{GB})]$ and fitted the observed impedance data at each temperature with this model. Here, $R_G$, $R_{GB}$ are the resistances of Gs and GBs, and $Q_G, Q_{GB}$ are the constant phase elements of Gs and GBs, respectively. Here, the constant phase element ($Q$) is used for the non-ideal capacitance and $C = R^{(1-n)/n} Q^{1/n}$ (where $n = 1$ for ideal capacitance). The observed data are fitted using the Eq. 13 & 14, considering that PZM consists of GBs and Gs.

$$Z' = Z'_{GB} + Z'_{G} = \frac{R_{GB}}{1+\left(\omega Q_{GB}^{\frac{1}{n_{GB}}} R_{GB}^{\frac{1}{n_{GB}}}\right)^2} + \frac{R_G}{1+\left(\omega Q_{G}^{\frac{1}{n_{G}}} R_{G}^{\frac{1}{n_{G}}}\right)^2} \tag{13}$$

and

$$Z'' = Z''_{GB} + Z''_{G} = \frac{\omega Q_{GB}^{\frac{1}{n_{GB}}} R_{GB}^{\frac{1+n_{GB}}{n_{GB}}}}{1+\left(\omega Q_{GB}^{\frac{1}{n_{GB}}} R_{GB}^{\frac{1}{n_{GB}}}\right)^2} + \frac{\omega Q_{G}^{\frac{1}{n_{G}}} R_{G}^{\frac{1+n_{G}}{n_{G}}}}{1+\left(\omega Q_{G}^{\frac{1}{n_{G}}} R_{G}^{\frac{1}{n_{G}}}\right)^2} \tag{14}$$

where, $Z'$ and $Z''$ are the real and imaginary parts of impedance, respectively. The subscript "GB" and "G" are indicating the contribution of GBs and Gs respectively, $\omega$ is the angular frequency, $n$, $Q$ and $R$ are the exponent, constant phase element and the resistance, respectively.

The black solid lines in FIG. 6 (a) & 6 (b) represent the observed data fitted using the electrical equivalent circuit. FIG. 6 (c) & inset of FIG. 6 (c) show the fitted $R_{GB}$ and $R_G$ data against the temperature. These two figures show that $R_{GB}$ and $R_G$ values decrease with the increase in temperature, which signifies the semiconductor-like behavior of PZM. Similarly, FIG. 6 (d) & inset of FIG. 6 (d) show the variation of $n_{GB}$ and $n_G$ with the temperature, respectively. It also is seen that the $n_{GB}$ and $n_G$ values increase with the increase in temperature. The value of $n_{GB}$ varies from 0.956 at 300 K to 0.975 at 700 K, whereas, the value of $n_G$ varies from 0.77 at 300 K to 0.845 at 700 K. This suggests that both GB capacitance and G capacitance move towards the ideal behavior (where $n = 1$). These results indicate the reduction of defects at GBs as well as in grain interior with increasing temperature. But, the value of $n_{\mathrm{G}}$ (0.845) at 700 K is also suggests that there are still some electronic or ionic defects.

To correlate the conduction mechanism through the impedance spectroscopy, we have plotted the previous fitting parameters $R_{\mathrm{G}}$ and $R_{\mathrm{GB}}$ using the NNH model using ~~the~~ following Eq. 15 as shown in FIG. 7 (a) & 7 (b), respectively.

$$R/T = R_{\alpha} e^{\left(E_a/k_B T\right)} \quad (15)$$

where $R_{\alpha}$ is the pre-exponential term, $E_a$ is the activation energy, and $k_B$ is the Boltzmann constant. The linear trends in FIG. 7 (a) & 7 (b) show a reasonably good fit to the NNH model above 580 K with corresponding activation energies 0.15 eV and 0.18 eV for the Gs and GBs, respectively. Below 580 K, the trends show non-linear behavior, which suggests the existence of another conduction model in PZM. In this region, the charge carriers prefer to hop between the sites lying within a certain range of energies. By employing the MVRH model, the impedance data are well fitted using the following relation:

$$\ln(R/R_0) = (T/T_0)^{1/4} \quad (16)$$

where $R_0$ is the pre-exponential factor and $T_0$ is the characteristic temperature. Insets of FIG. 7 (a) & 7 (b) show a good agreement between the observed data and the MVRH model.

## F. Complex permittivity & modulus

FIG. 8 (a) & 8 (b) show the frequency dependence behavior of dielectric constant (ε') & loss tangent (tanδ) for PZM at various temperatures, respectively. It is observed that the value of ε' decreases with increasing frequency. In the measured frequency range, two dielectric dispersions in ε' and corresponding two peaks (one peak at low frequency and another peak at high frequency) in tanδ are observed, which suggests the existence of two different types of the relaxation processes in PZM. At low frequency, the dielectric dispersion and the corresponding peak at $\tan\delta$ are due to the existence of GBs in the system, whereas at high frequency, that is due to the existence of Gs. These two dispersions in $\varepsilon'$ and the corresponding two peaks in the $\tan\delta$ move towards a high-frequency region with the increase of temperature, which suggests thermally activated relaxation process. In the measured frequency window, it also seen that the low frequency dispersion in the $\varepsilon'$ and the corresponding larger peak in $\tan\delta$ due to GBs contribution, which enters from the low-frequency side above 500 K, on the another side, the high frequency dispersion in $\varepsilon'$ and corresponding smaller peak in $\tan\delta$ originated due to the Gs contribution, which gradually shift towards the high frequency side with increasing temperature. The maximum value of $\tan\delta$ is found 0.4 – 0.5 for GBs and 0.2 – 0.25 for Gs, as described by other researchers for similar type systems [24, 50].

The variation of $\varepsilon'$ with the temperature at various frequencies is shown in FIG. 8 (c). This spectrum shows the increase in $\varepsilon'$ with the temperature, which shows no relaxor-like behavior. As the number of charge carriers increases with increasing temperature, the polarization inside the material increases, which results increase~~s~~ in $\varepsilon'$. With further increase in temperature, the material exhibits colossal dielectric constant behavior, which is similar to the literature of $CaCu_3Ti_4O_{12}$ [51]. However, in the measured temperature range, the increase in $\varepsilon'$ is a step-like curve, which shifts towards the high temperature side with the increasing frequency. This type of behavior is observed due to the existence of the two different types of relaxation mechanisms, which can be described by the two parallel RC circuits connected in series with different RC values as described in Sec. III (E). Here the values of C and R for GBs are higher than that of Gs, which creates a high resistive thin layer that effects in $\varepsilon'$. This high capacitive thin layer offering strong resistance to the flow of the charges, which results in high $\varepsilon'$ at low frequency and high temperature. The dependence of $\tan\delta$ on temperature is shown in FIG. 8 (d). This graph shows two sets of peaks in tanδ. With the increasing frequency, the positions of the observed two sets of peaks shift towards high temperature, which suggesting the existence of two thermally activated relaxations in PZM. The tanδ peaks depend on the charge carriers' mobility and temperature. The charge carriers' mobility increases with the increase in temperature, and the charge carriers commence relaxing at the comparatively high-temperature side. This result, the tanδ peaks move towards the high-temperature side with increasing frequency.

The observed values of relaxation times ($\tau = 1/\omega_m$) are calculated from the tanδ peak's position (i.e., $\omega_m$, the maximum value of angular frequency at tanδ peak) from FIG. 8 (b). The variation of the relaxation times related to the GBs and the Gs are shown in FIG. 9 (a) & 9 (b) by utilizing the NNH model with Eq. 17.

$$\tau = \tau_0 e^{\left(E_a/k_B T\right)} \quad (17)$$

$$\tau = \tau_\alpha e^{\left(T_0/T\right)^{1/4}} \quad (18)$$

here, $\tau_0$ and $\tau_\alpha$ are the pre-exponential constants, $E_a$ is the activation energy, $k_B$ is the Boltzmann constant, $T_0$ is the characteristics temperature and $T$ is the absolute temperature. It is observed that the temperature dependence of $\tau_G$ (FIG. 9 (a)) & $\tau_{GB}$ (FIG. 9 (b)) follow the NNH model using the Eq. 17 for temperatures above 580 K. The values of $E_a$ are found to be 0.1487 eV and 0.1527 eV for Gs and GBs, respectively. These figures indicate the change in the relaxation mechanism at 580 K. The values of $\tau_G$ are plotted using the MVRH model (Eq. 18) is shown in FIG. 9 (c). It is also found a good deal between the observed data and the MVRH model. From these results, we can summarize that similar charge carriers are responsible for the conduction, impedance, and relaxation processes.

The electrical modulus is an important parameter, which provides us a piece of useful information about the charge transport dynamics [52]. FIG. 10 (a) & 10 (b) show the frequency dependence plot of real (M') & imaginary (M") parts of electric modulus (M*), respectively. It is seen that the M' value increases with increasing frequency and two step-like increases are found in the measured frequency range. These two step-like increase in M' corresponding two the peaks display in M". This behavior suggests the existence of two types of polarization effects. The first peak in M" and the corresponding first step-like increase in M' at low frequency is due to GBs contribution and similarly the second peak in M" and the corresponding second step-like increase in M' at high frequency is the effect of Gs contribution. The M' value decreases with the increase in temperature, and the corresponding peaks in M" shift towards the high-frequency range, which also suggesting thermally activated process. The low frequency ($< 10^2$ Hz) peak in M" suggests the charge carriers undergo long range hopping from the one site to their neighboring sites. On the other side, the peak in M" at high frequency originates due to the charge carriers' hopping, which is confined to the trapping center, and their mobility is restricted to short distances. The frequency dependence of M' can be explained by the CPE model:

$$M' = \omega C_0 Z'' = C_0 \left[ \frac{\omega^2 Q_{\mathrm{GB}}^{\frac{1}{n_{\mathrm{GB}}}} R_{\mathrm{GB}}^{\frac{1+n_{\mathrm{GB}}}{n_{\mathrm{GB}}}}}{1+\left( \omega Q_{\mathrm{GB}}^{\frac{1}{n_{\mathrm{GB}}}} R_{\mathrm{GB}}^{\frac{1}{n_{\mathrm{GB}}}} \right)^2} + \frac{\omega^2 Q_{\mathrm{G}}^{\frac{1}{n_{\mathrm{G}}}} R_{\mathrm{G}}^{\frac{1+n_{\mathrm{G}}}{n_{\mathrm{G}}}}}{1+\left( \omega Q_{\mathrm{G}}^{\frac{1}{n_{\mathrm{G}}}} R_{\mathrm{G}}^{\frac{1}{n_{\mathrm{G}}}} \right)^2} \right]$$

(19)

Here, $C_0$ is the capacitance of empty cells (i.e., without material PZM) and the values of $Z''$ are collected from Sec. III (E). It is found that this CPE model fits observed data successfully in FIG. 10 (a).

## G. Electrochemical activity

FIG. 11 (a) shows the CV curves of PZM at various scan rates 2, 5, 10, 20, 50, and 100 mV/s within a potential window -0.8 to 0.5 V (vs Ag/AgCl) in 1 M aqueous KOH electrolyte at room temperature. One distinct oxidation peak (~ 0 V vs Ag/AgCl) and one reduction peak (~ -0.45 V vs Ag/AgCl) are observed in each curve which suggests the pseudocapacitive behavior of PZM due to its nonrectangular CV loops. This is quite different from the conventional electrical double-layer capacitors (DLC). The potential window over, which the redox peaks are observed is consistent with the prior studies of Mn-based electrodes [15]. The existence of oxidation and reduction peaks in the CV curves can ~~be~~ classified due to the existence of the Faradic redox reaction of Mn between the 4+ and 3+ oxidation states at the surface or inside the bulk. The oxidation peak at 0 V corresponds to $Mn^{3+} \rightarrow Mn^{4+}$ as the oxygen vacancy sites are filled, whereas the reduction peak at -0.45 V corresponds to the opposite reaction of the oxidation into the material. A redox peak pair exhibits the deintercalation/intercalation of oxygen from/into pseudocapacitor electrodes. It is also found that the electrode's current response increases with increasing the scan rates and the oxidation & reduction peaks shift towards the positive & negative potential, respectively, which indicating a quasi-reversibility of the electrode. This quasi-reversibility nature indicates the Faradic redox reaction dominates the specific capacitance.

In an anion based intercalated pseudocapacitor, the diffusion of $O^{2-}$ anion in the bulk oxide and the absorption of $OH^-$ anion on the electrode surface are the two main leading processes [53, 54], which can be described using the Power law [55]:

$$i_p = a\nu^b \qquad (20)$$

Here $i_p$ is the peak current, $a$ and $b$ are the adjustable parameters, and $\nu$ be the scan rate (mV/s). The values of $b$ lie between 0 and 1. The value of $b$ is 0.5 signifies that the current is run by a semi-infinite diffusion process. On the other hand, the $b$ value of unity implies the surface controlled current and the charge storage process is purely capacitive. The plot of log $i_p$ vs log $\nu$ for the cathodic peak is shown in FIG. 11 (b). The $b$ value of the cathodic peak is found to be 0.523, suggesting that the current comes primarily from the oxygen ion intercalation diffusion process in the PZM electrode. The peak current ($i_p$) depends on the scan rates ($\nu$) according to the Randles-Sevcik equation [56]:

$$i_p = 0.4463\, nFAC \left(\frac{\alpha nFD}{RT}\right)^{1/2} \nu^{1/2} \quad (21)$$

or if the solution is at $T = 25°C = 298$ K, the above equation becomes:

$$i_p = 268600 n^{\frac{3}{2}} A D^{\frac{1}{2}} C \nu^{\frac{1}{2}} \quad (22)$$

where $n$ is the number of electrons transferred in the redox reaction, $F$ is the Faraday constant, $A$ is the surface area of the electrode, $C$ is the maximum concentration in mol/cm$^3$, $D$ is the diffusion coefficient, $\alpha$ is the transfer coefficient, $R$ is the ideal gas constant and $T$ is the absolute temperature. The value of $D$ is found to be $1.576 \times 10^{-5}$ m$^2$/s for the bulk PZM electrode. However, to clarify the contribution of the surface capacitive process in PZM electrode, the $i_p$ can be expressed in the combination of two processes:

$$i_p = k_1 \nu + k_2 \nu^{1/2} \quad (23)$$

where $k_1$ and $k_2$ are two constants. The first term $k_1 \nu$ and the second term $k_2 \nu^{1/2}$ correspond to the $i_p$ contributions from the surface capacitive process and the diffusion-controlled intercalation process, respectively. To calculate the values of $k_1$ and $k_2$, we rearrange the Eq. 23 into Eq. 24:

$$i_p/\nu^{1/2} = k_1 \nu^{1/2} + k_2 \quad (24)$$

Inset of FIG. 11 (b) shows $i_p/\nu^{1/2}$ vs $\nu^{1/2}$ plot for cathodic peaks. The plot shows two distinct regimes: regime I (at low scan rates, $\nu$ = 2 to 10 mV/s) and regime II (at high scan rates, $\nu$ = 20 to 100 mV/s). In regime I, the values of $k_1$ and $k_2$ are found to be 0.0170 (25.87%) and 0.0487 (74.13%), respectively. This suggests a major contribution from the oxygen ion diffusion process inside the electrode and minor contribution from the surface capacitive process in the PZM. On the other hand, in regime II, the values of $k_1$ and $k_2$ are found to be 0.0004 (0.5%) and 0.0817 (99.5%), respectively, which implies an oxygen ion diffusion process is dominated in this regime. When the scan rate increases, all the interfaces of the electrode not getting sufficient time to experience the redox reactions. Consequently, the Spartan kinetics limit the charge transfer process at the electrode-electrolyte interface, and the diffusion process achieves a very high charge storage level with no delay in time.

FIG. 11 (c) shows the GCD curves at various current densities (2, 3, 4, 5, 6, 8, and 10 A/g) for the same potential window of CV measurement. The value of $C_{sp}$ at various current densities can be calculated from the Eq. 25:

$$C_{sp} = I\Delta t/Vm \tag{25}$$

where $I$ is the discharge current, $\Delta t$ is the discharge time, $V$ is the measured potential window, and $m$ is the active mass of PZM. The value of $C_{sp}$ is found to be 69.14 F/g ($\times 10^3$ F/kg) at a current density of 2 A/g ($\times 10^3$ A/kg), which is quite similar to the previously reported literature value for the bulk perovskite electrodes [57]. The $C_{sp}$ remains at 48.93 F/g, when current density becomes 10 A/g (Inset of FIG. 11 (c)) which suggests 70% retention when a 5-fold increase in the current density. The long period cyclic performance of the bulk PZM electrode is investigated up to 500 cycles in 1 M aqueous KOH electrolyte at 3 A/g current density, as displayed in FIG. 11 (d). After 500 cycles, the $C_{sp}$ of the PZM electrode reduces to 94% suggests the good stability of the electrodes. Here, it is also found that during the long-term cyclic test, the charge/discharge profile remains the same type of symmetric triangular shape. This implies a stable electrochemical performance and no structural change during a charge transfer reaction in the PZM electrode. The energy density (E) was calculated using the formula $\frac{1}{2}C_{sp}V^2$ and it is found that the value of E decreases from 16.23 $Whkg^{-1}$ at 2 A/g to 11.48 $Whkg^{-1}$ at 10 A/g. On the other hand, the power density (P) was calculated using $E/\Delta t$ and its value increases from 1.2268 $kWkg^{-1}$ at 2 A/g to 14.251 $kWkg^{-1}$ at 10 A/g. Inset of FIG. 11 (d) shows the EIS measurement of the bulk PZM which is carried out at an applied potential of 0.4 V within a frequency range from 100 mHz to 100 kHz with an ac voltage amplitude of 5 mV. The plot clearly shows the inclination with the X-axis at low frequency range, whereas parallel to the X-axis at high frequency range, which suggests that the charge transfer rate is faster at high frequency range than that at the low frequency range. The GBs and the other inter-particle phenomena may affect the outer surface capacitance. The inclination at the lower frequency range indicates the frequency dependent linear diffusion process, commonly known as Warburg impedance ($Z_w$), which is due to the high diffusion rate of ions during the redox reactions. At the high frequency range, the intercept of the plot on the X-axis represents the equivalent series resistance ($R_s$) and the semicircle signifies the charge transfer resistance ($R_{ct}$) of the electrode-electrolyte interface. The calculated values of $R_s$ and $R_{ct}$ are found to be 1.06 Ω and 2.64 Ω, respectively. These two low values of $R_s$ and $R_{ct}$ for the bulk PZM suggesting higher electrical conductivity and the fast ion-charge transfer during the redox reaction.

The oxygen non-stoichiometry δ of the PZM is calculated in Sec. III (C) and the value is 0.24 as the oxygen vacancy is about 4% in the material. The pseudocapacitive behavior of the bulk PZM is characterized using the CV curve in 1 M aqueous KOH solution. In bulk PZM electrode, the cation Mn forms Mn–O bond, which provides channels for the oxygen ions to migrate. This suggests that the PZM has flawless oxygen-ion based conductivities. The oxygen intercalation and discharge processes in the bulk PZM electrode can be written as:

$$Pr_2Zn[Mn_{2\delta}^{3+}, Mn_{(1-2\delta)}^{4+}]O_{6-\delta} + 2\delta OH^- \underset{discharge}{\overset{charge}{\rightleftarrows}} Pr_2ZnMn^{4+}O_6 + \delta H_2O + 2\delta e^- \quad (26)$$

FIG. 12 shows the above charge transfer process in the bulk PZM electrode. During charging, oxygen vacancies are occupied by the oxygen ion intercalation and diffusion of oxygen ion in the lattice, accompanied by the oxidation of $Mn^{3+}$ to $Mn^{4+}$ (FIG. 12 (a)), whereas during discharging, the reaction is reversed (FIG. 12 (b)). This charge-discharge process is similar to the $LaMnO_{3\pm\delta}$ [15] and $LaNiO_{3-\delta}$ [58] electrodes. These two works of literature confirm that the charge storage mechanism in perovskite based pseudocapacitor electrodes is mutually related to oxygen transport and change in the valence of B-cation. This investigation will provide a piece of good information for the more improvement of the stable and high-performance perovskite-based anion-intercalated electrode material.

## IV. CONCLUSION

A combined study of the XRD and Raman spectrum of PZM authenticate the monoclinic $P2_1/n$ structure. The existence of a super-lattice peak (101) in the XRD pattern indicates the corresponding tilting of $MnO_6$ octahedra for monoclinic symmetry, respectively. Also, $A_g$ mode reveals the breathing vibration of the $MnO_6$ octahedra in $P2_1/n$ phases. The X-ray photoemission spectrum suggests the existence of oxygen vacancy in PZM, which plays a crucial role in electrical conduction as well as in electrochemical behavior. The Nyquist plot, complex permittivity, and electric modulus highlight the blended effect of Gs and GBs, which leads to the charge carrier dynamics in PZM. The existence of two various conduction mechanisms (NNH and MVRH) is observed using a combined study of dc conductivity, impedance, and complex permittivity in PZM. This investigation points towards a transition from NNH to MVRH mechanism due to decreasing activation energy around 580 K. The Nyquist plots and electric modulus are well fitted using the CPE model. The bulk PZM electrode exhibits intercalated pseudocapacitive nature and also offering $C_{sp}$ of 69.14 F/g at a charge and discharge currents of 2 A/g in 1 M aqueous KOH electrolyte solution with larger cycle life. This study will provide a good knowledge for the further development of stable and high-performance perovskite anion-intercalated electrode material for supercapacitor applications.

## SUPPLEMENTARY MATERIALS

See this section for Supplementary TABLEs ST1, ST2, ST3 and FIG. SF1.

## ACKNOWLEDGMENT

MR(NET JRF ID no. 522407)acknowledges the financial supports provided by the University Grants Commission (UGC), India.

## DATA AVAILABILITY STATEMENT

The data that support the findings of this study are available from the corresponding author upon reasonable request.

Figure

**Figures**

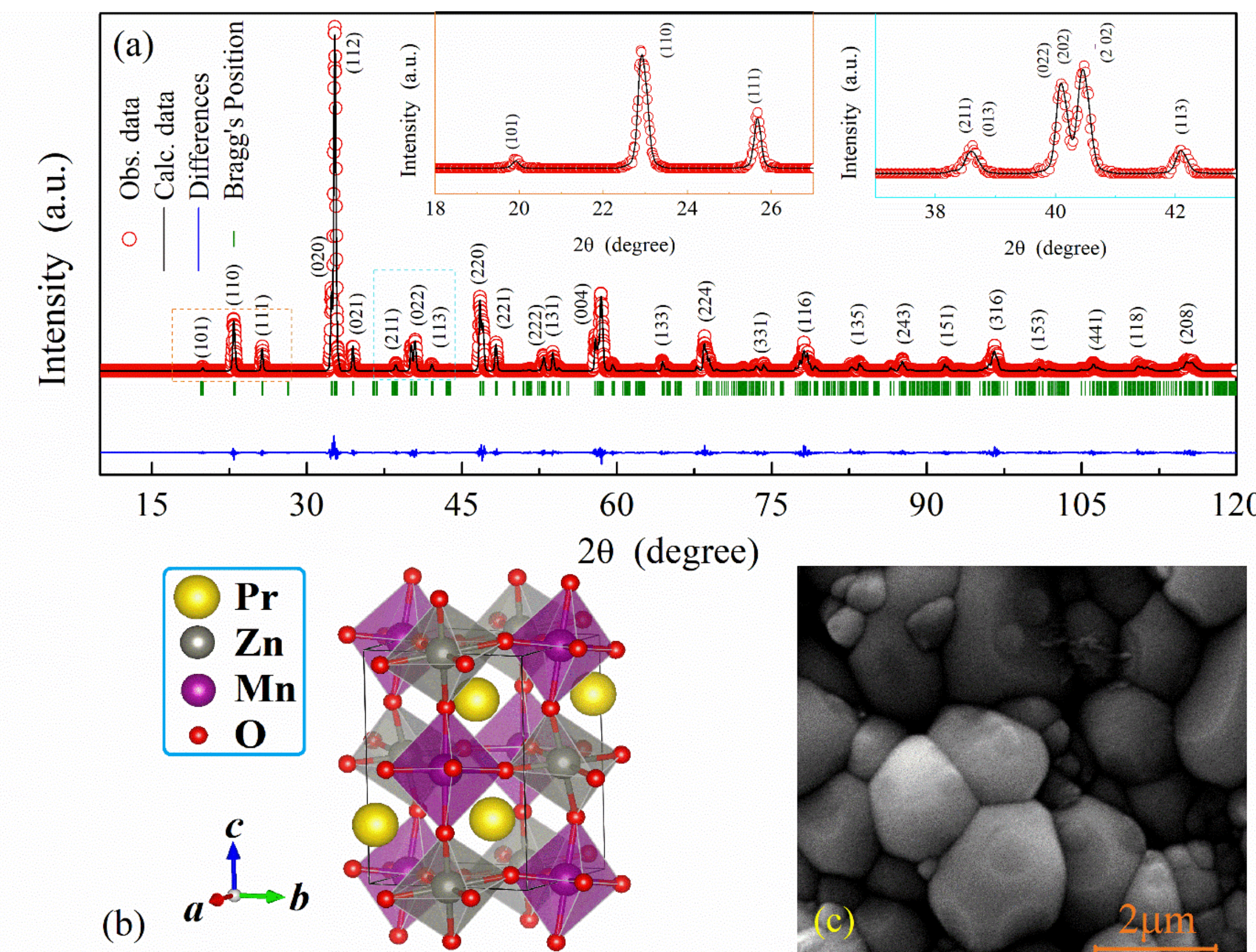

**FIG. 1**

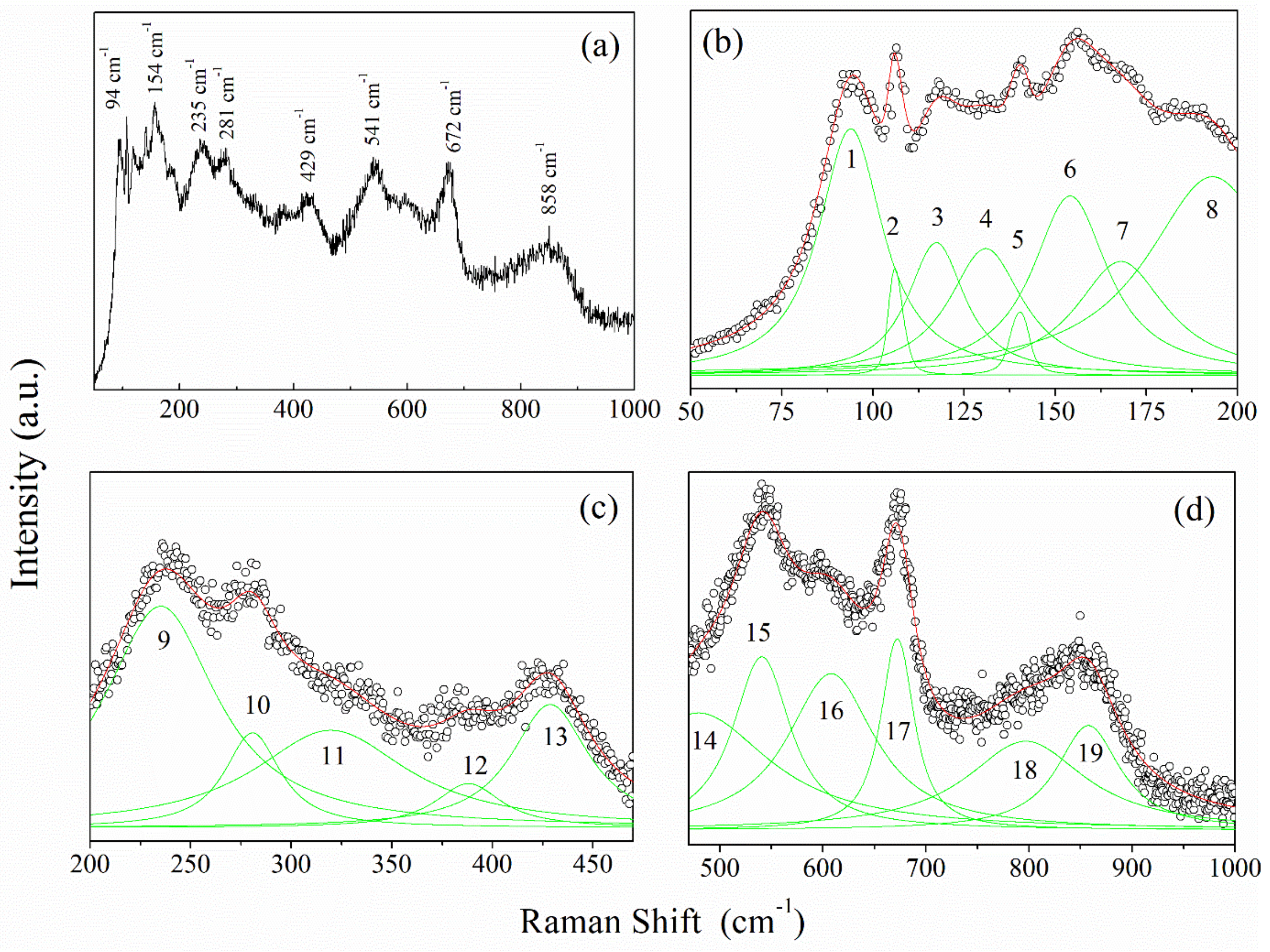

**FIG. 2**

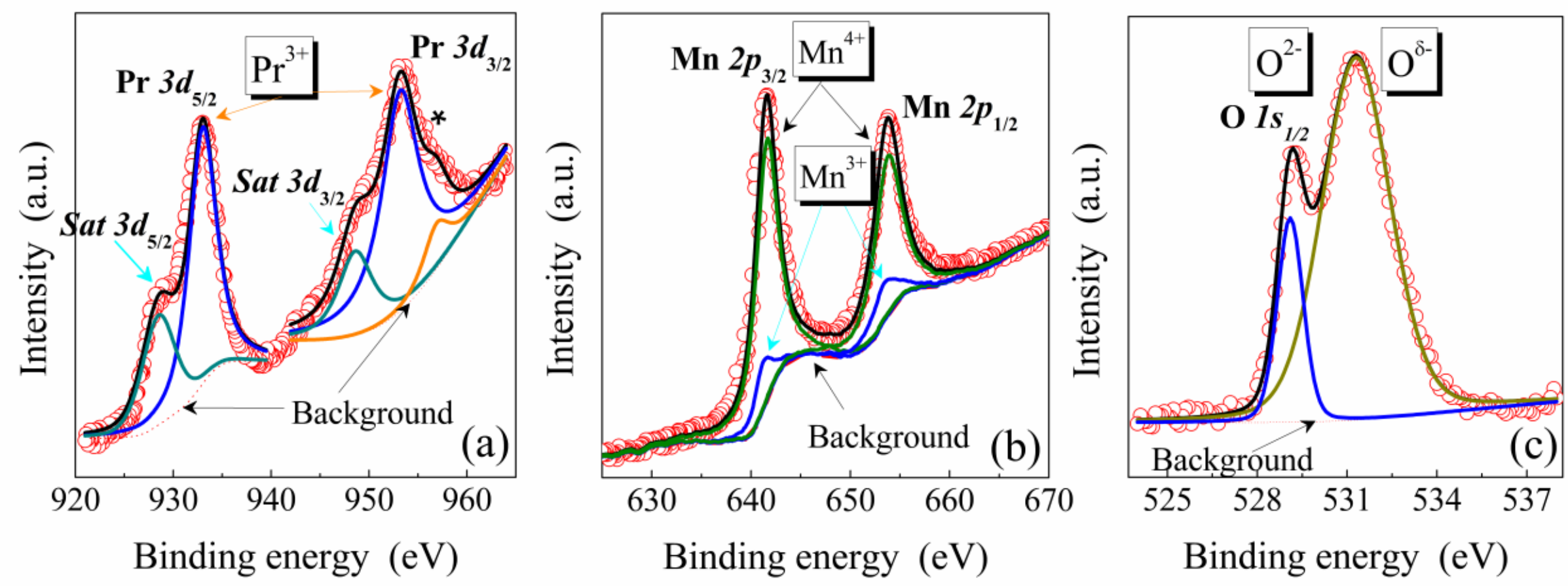

**FIG. 3**

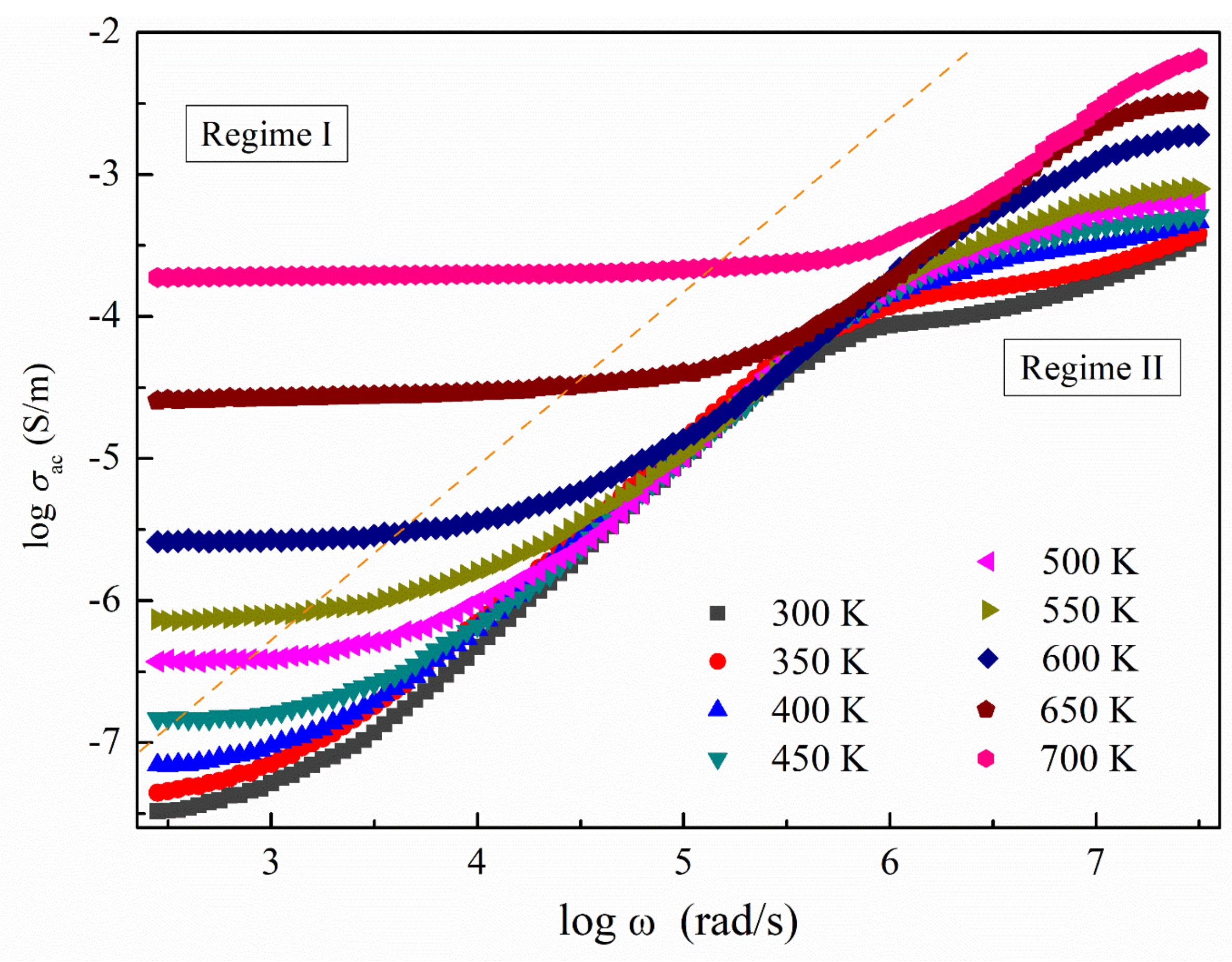

**FIG. 4**

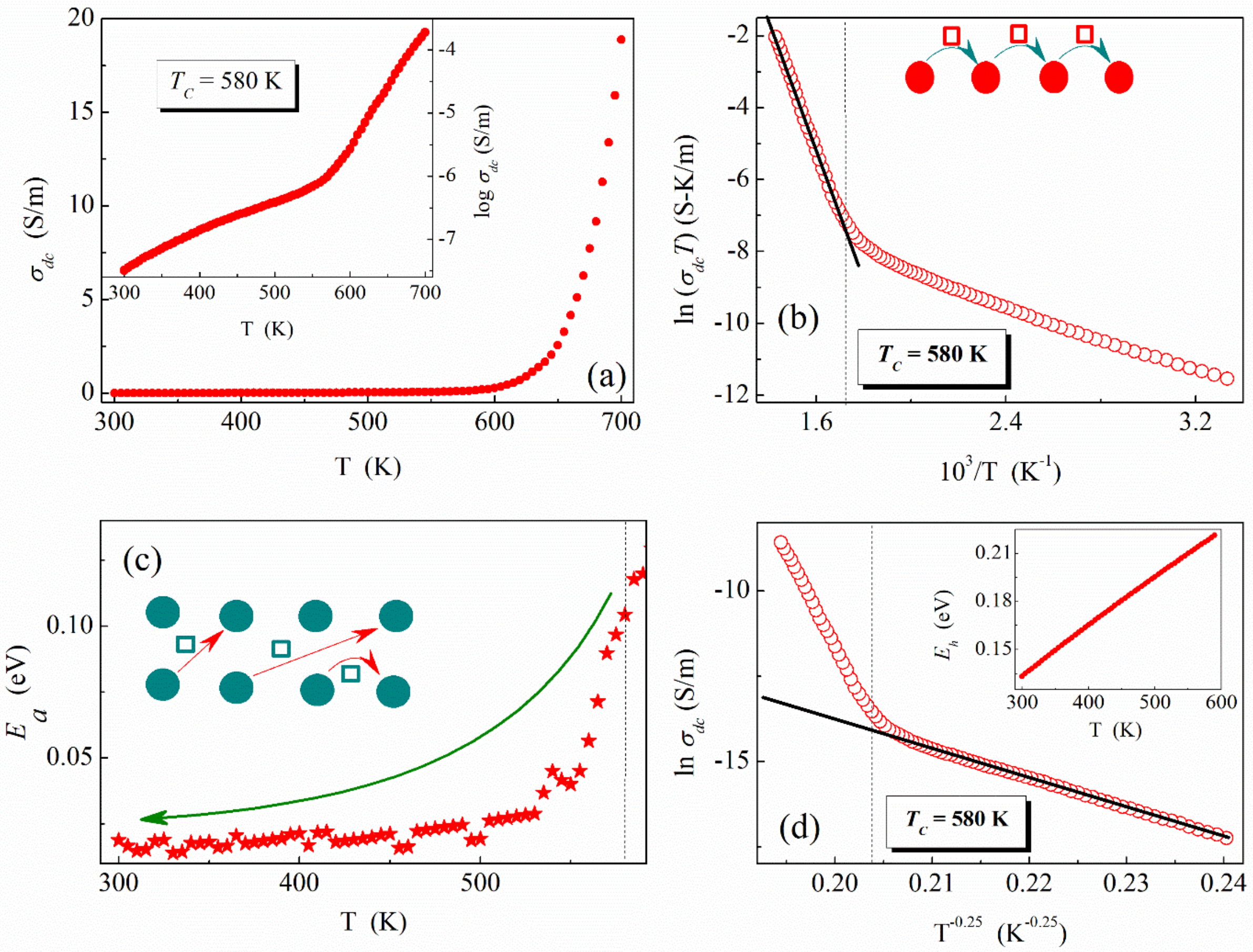

**FIG. 5**

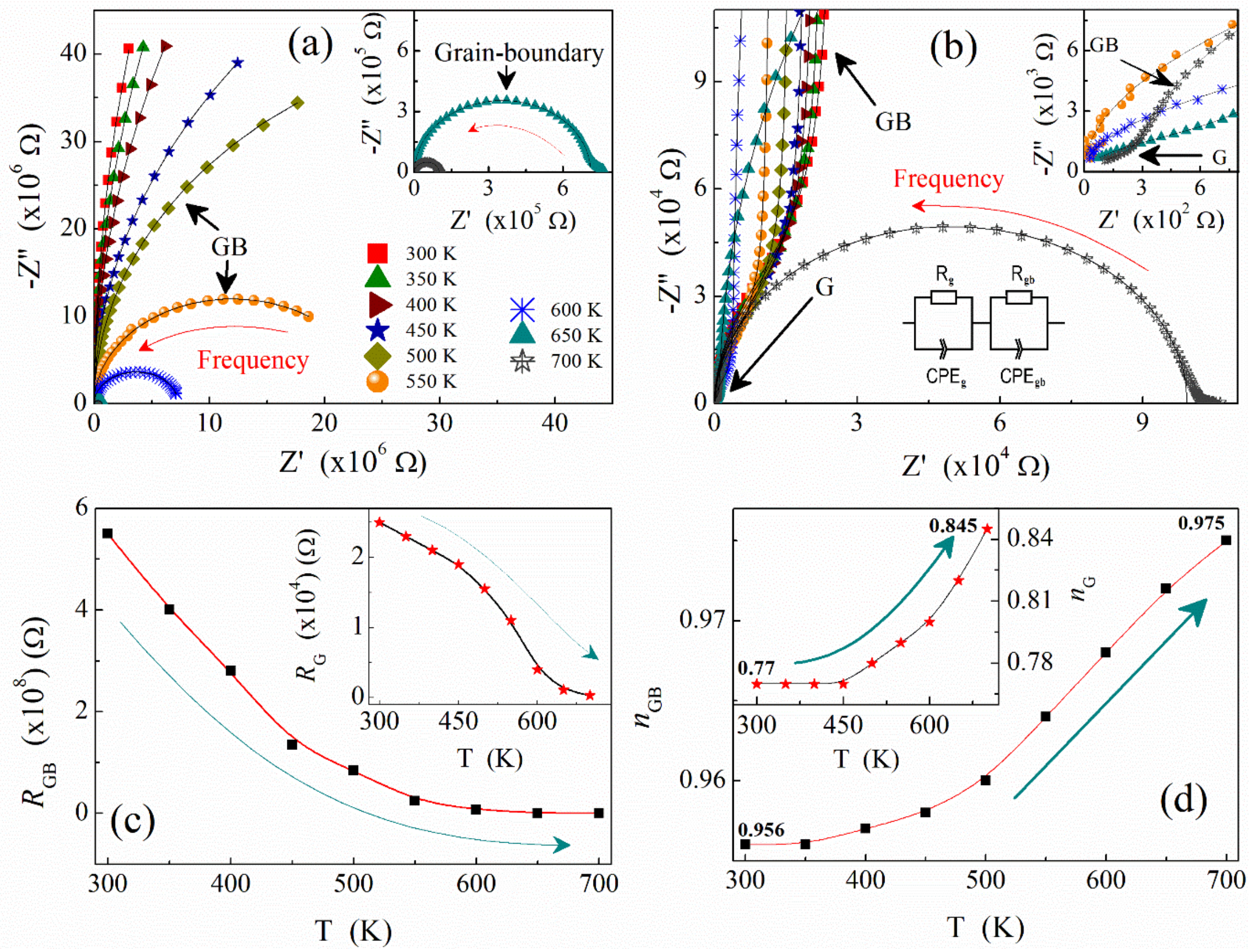

**FIG. 6**

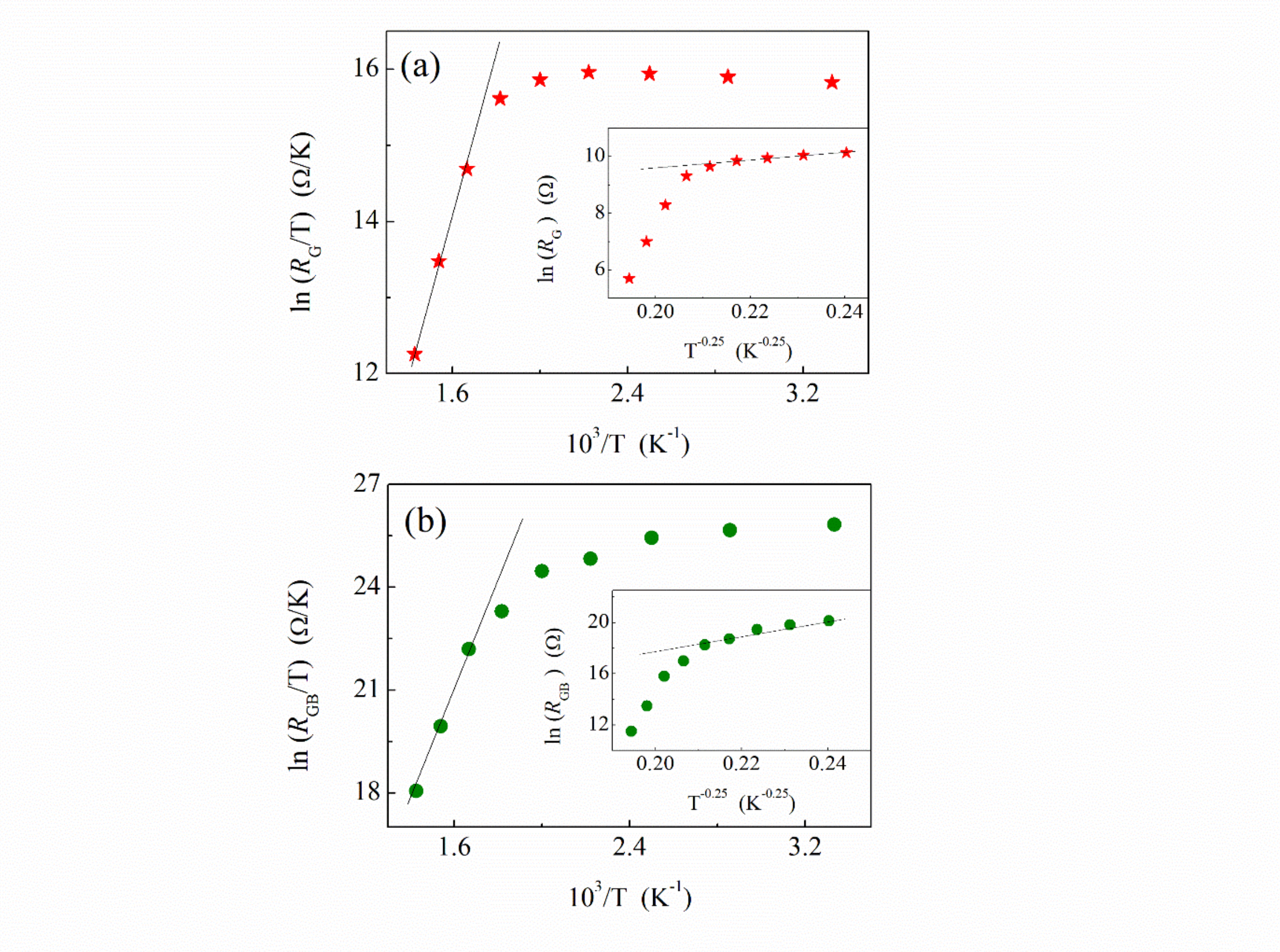

**FIG. 7**

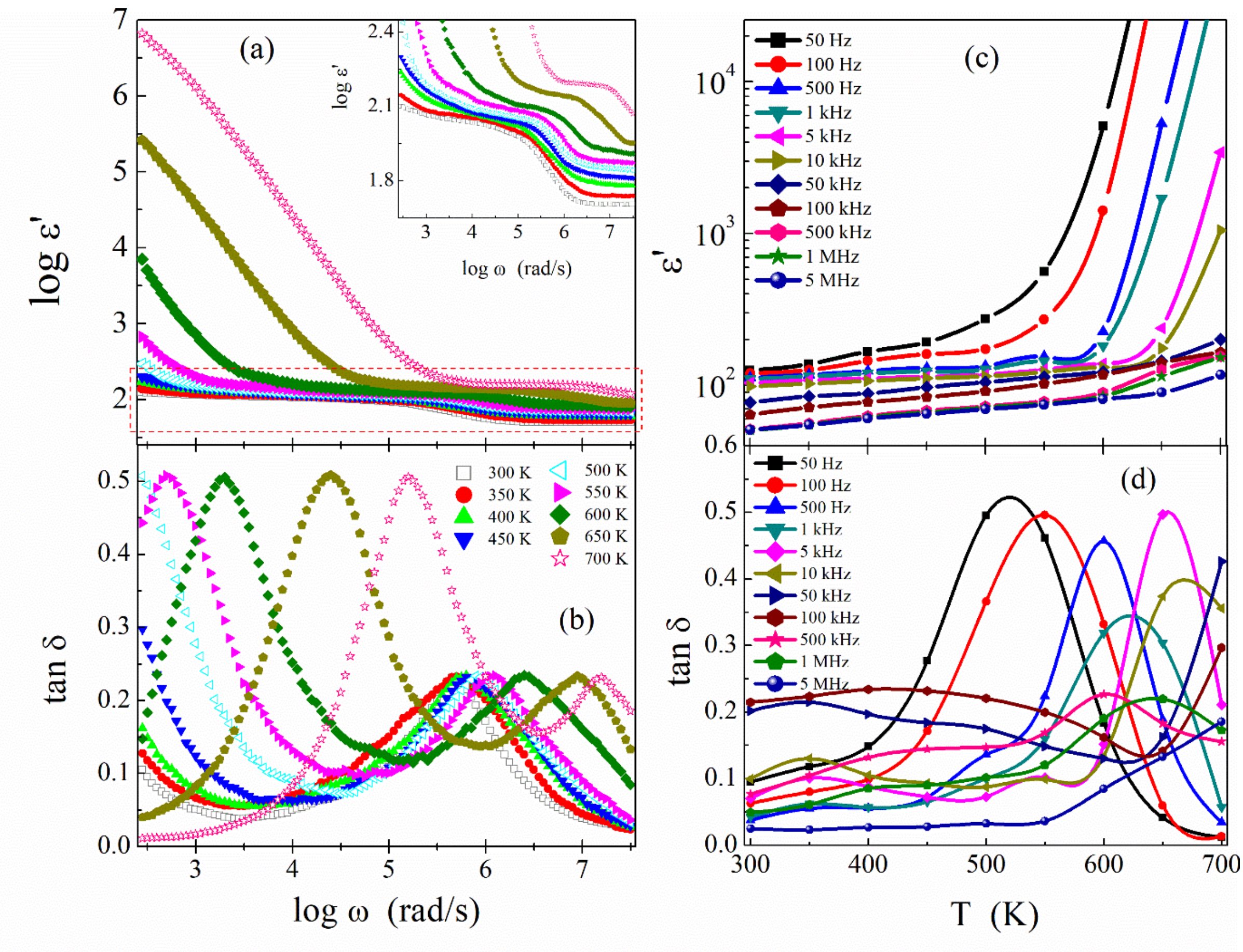

**FIG. 8**

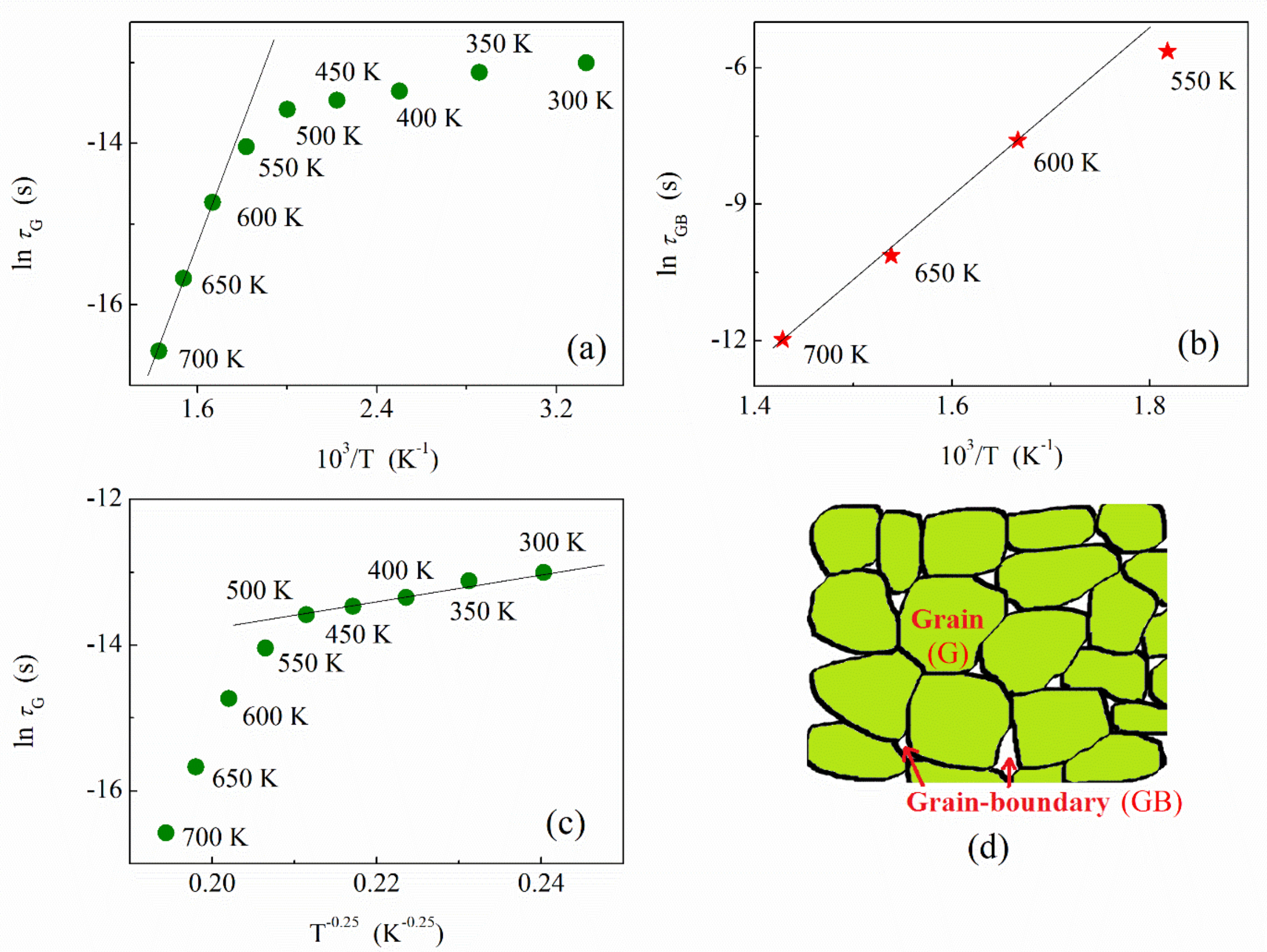

**FIG. 9**

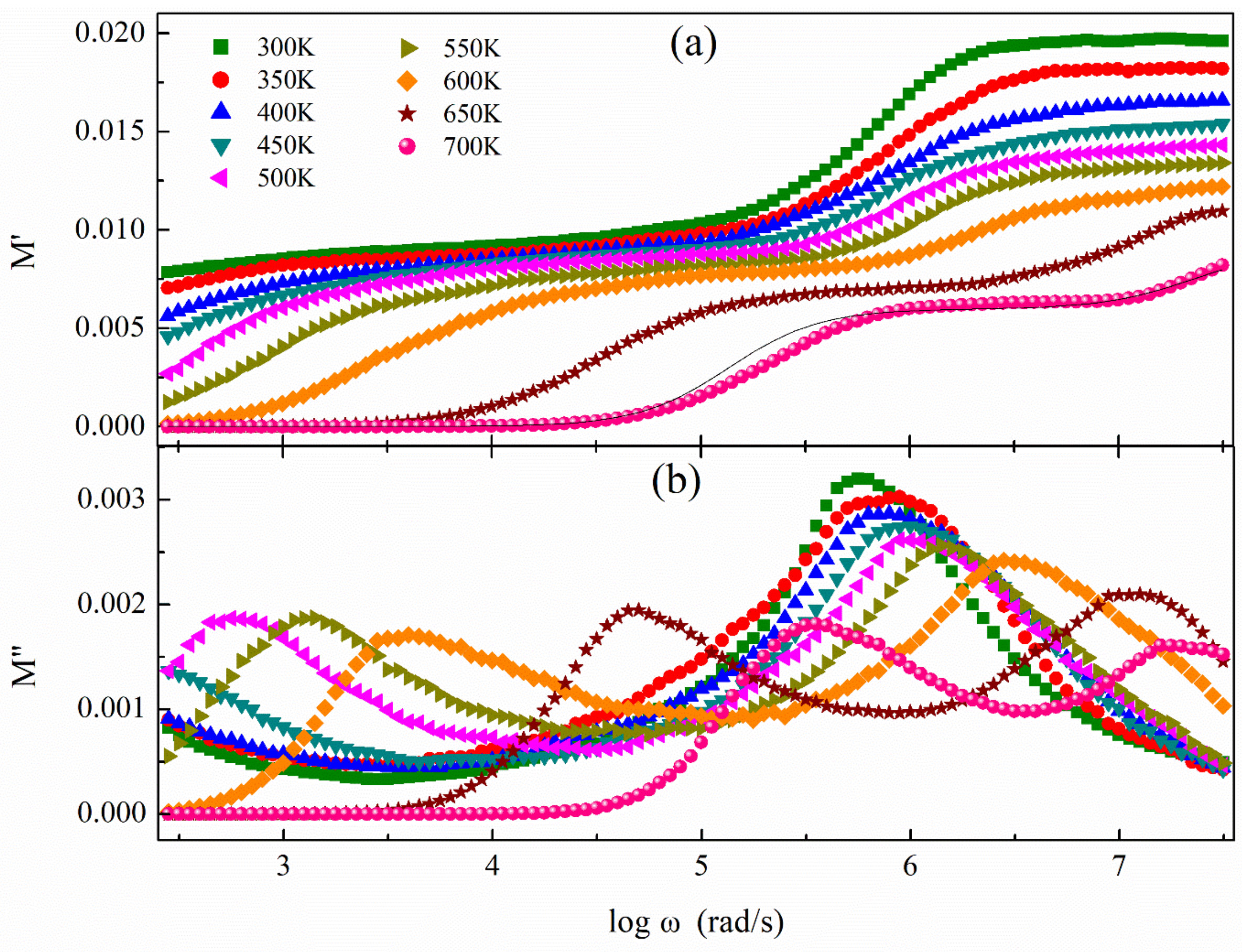

**FIG. 10**

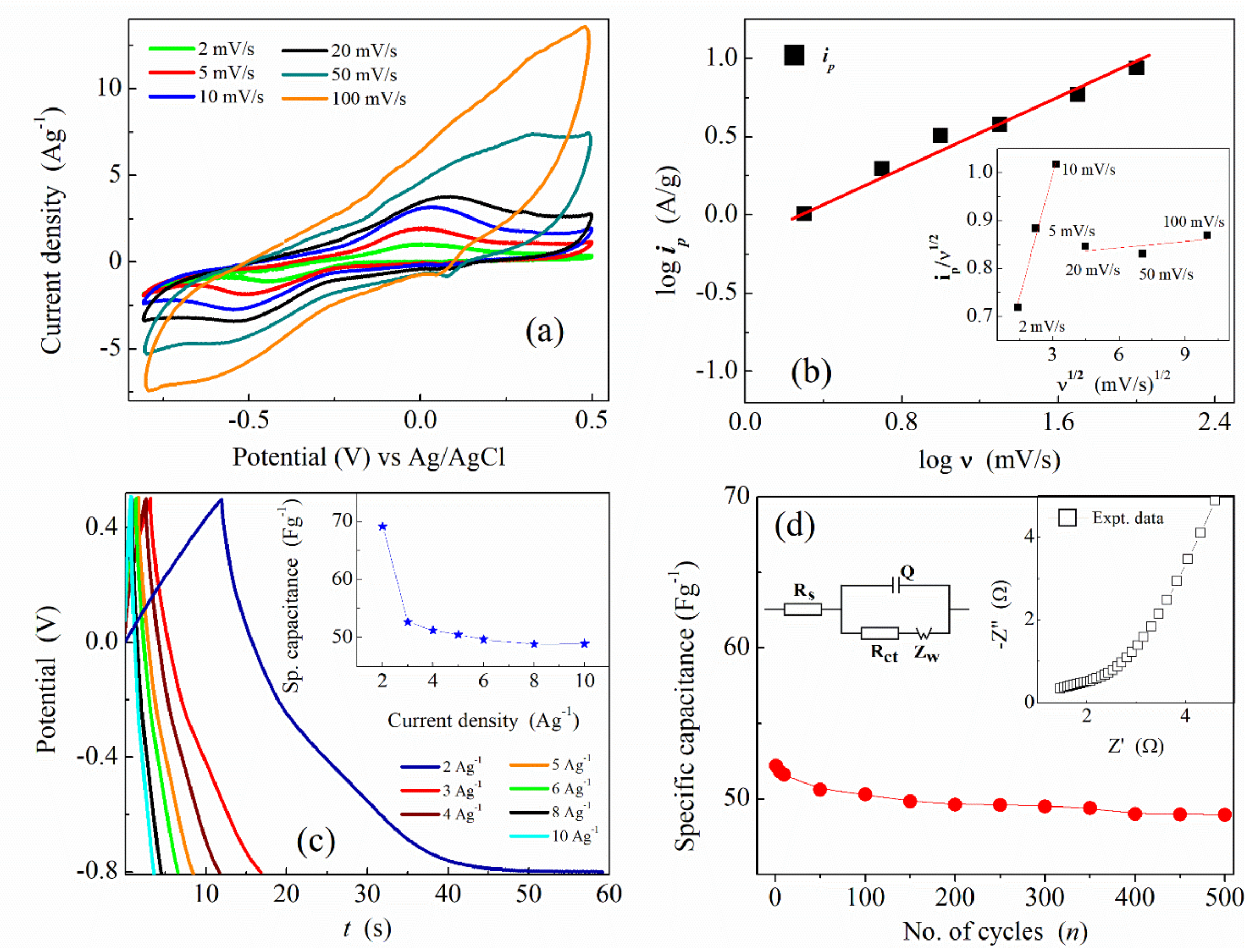

**FIG. 11**

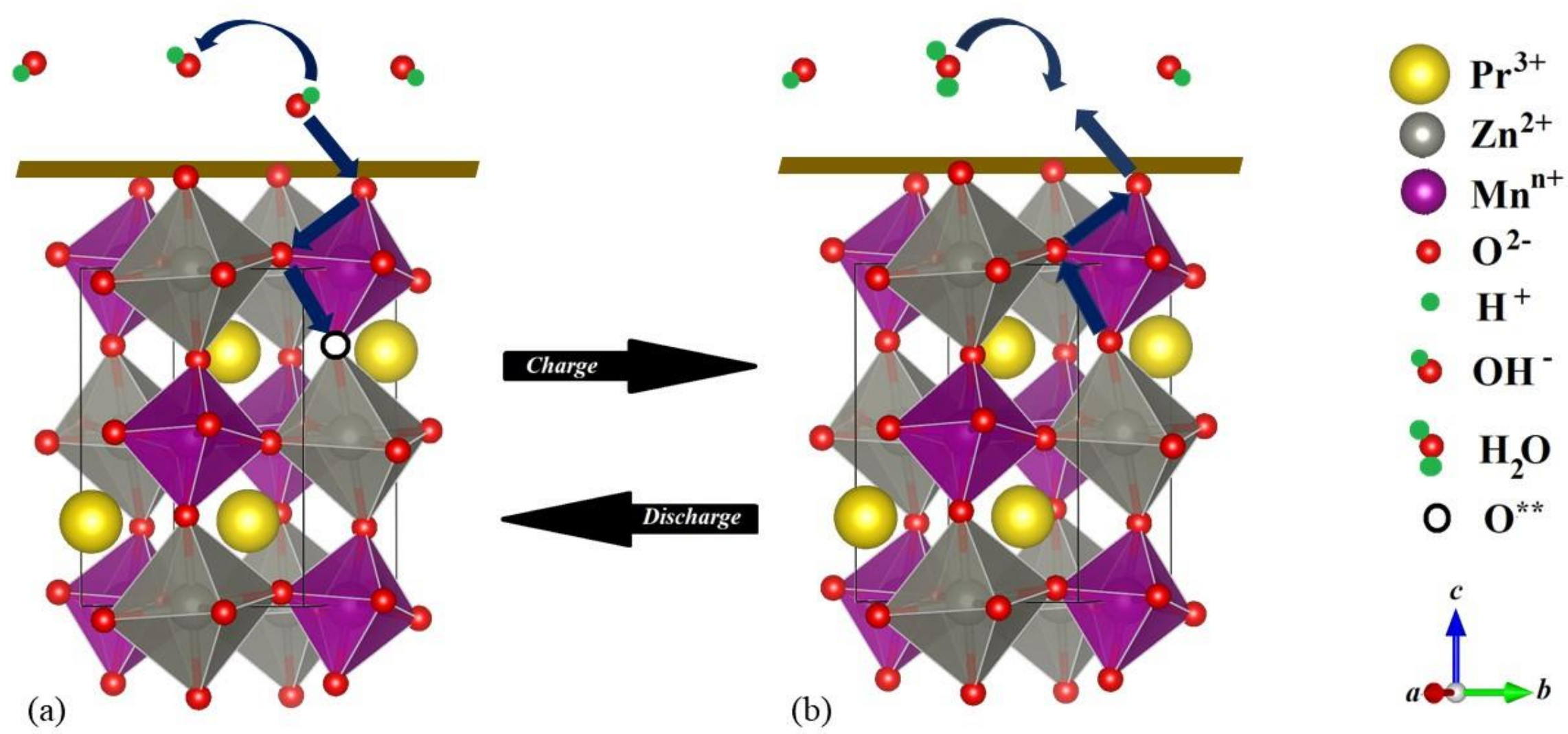

**FIG. 12**

**Figure captions**

FIG. 1: (a) The powder XRD profile for PZM at room temperature. (b)The unit cell of PZM crystallizes in monoclinic $P2_1/n$ as obtained from Rietveld refinement. (c) The SEM micrograph of PZM.

FIG.2: (a) The room temperature unpolarized Raman spectrum of PZM.The entiredeconvoluted Raman spectrum is separated for better view into three ranges: 50 – 200 $cm^{-1}$ (b), 200 – 470 $cm^{-1}$ (c) and 470 – 1000 $cm^{-1}$ ($\times 10^2$ $m^{-1}$) (d). (Observed data are open circles, while the solid lines represent phonon modes adjusted by Lorentzian curves).

FIG. 3: The XPS spectra of Pr*3d* (a), Mn *2p* (b) and O *1s* (c) regions for PZM.

FIG. 4: The variation of ac conductivity with frequency at various temperatures.

FIG. 5: (a) Thevariation of dc conductivity as a function of temperature for PZM. Inset shows logarithmic of dc conductivity with the temperature plot for clear view. The dc conductivity of PZM describes the conduction process using (b) nearest-neighbor hopping model, (d) Mott's variable range hopping model. The insets show the variation of hopping energy (d) as a function of temperature; (c) Variation of activation energy with the temperature below 590 K.

FIG. 6: Nyquist plots at different temperatures for PZM (a, b). Solid lines are fitted data to the observed data (points). Inset (b) shows the equivalent circuit used to fit the observed data. Grain-boundary resistance ($R_{GB}$) (c) and $n_{GB}$(d) are plotted with the temperature. Insets show the thermal variation of $R_G$ (c) and $n_G$(d).

FIG. 7: The grain ($R_G$) and grain-boundary ($R_{GB}$) resistances are plotted using (a, b) nearest-neighbor hopping model and (Insets of (a, b)) Mott's variable range hopping model.

FIG. 8: Frequency dependence of $\varepsilon'$ (a) and $\tan\delta$ (c) at various temperatures for PZM; Inset (a) shows the closed view of $\varepsilon'$ with frequency. The variation of $\varepsilon'$ (b) and $\tan\delta$ (d) with temperature at various frequencies for PZM.

FIG. 9: Relaxation times $\tau_{GB}$ and $\tau_G$ of the carriers at grain-boundaries and grains are plotted using (a, b) nearest-neighbor hopping model. The values of $\tau_{GB}$ are plotted using (c) Mott's variable range hopping model. (d) The figure illustrates the grains and grain-boundaries in PZM sample.

FIG. 10: Frequency dependence of the M′ (a) and M″ (b) of PZM at various temperatures. Solid black line is the fitting to the observed data at 700 K using CPE model.

FIG. 11: (a) The CV curves of PZM electrode at various scan rates, (b) log $\boldsymbol{i_p}$ vs log $\boldsymbol{\nu}$ plot and inset shows $\boldsymbol{i_p}/\boldsymbol{\nu}^{1/2}$ vs $\boldsymbol{\nu}^{1/2}$ plot, (c) The GCDcurves of PZM electrode at various current density, inset of (c) shows the variation of specific capacitance with current density, (d) Cycling performance of PZM at a current density 3 A/g ($\times 10^3$ A/kg), Insets of (d) show the Nyquist plot and the respective equivalent circuit used.

FIG. 12: Schematic diagram of (a) oxygen intercalation into PZM and (b) oxygen deintercalation from PZM.

**Tables:**

**TABLE 1**

Aristotype Structure

Cubic $\boldsymbol{Fm\bar{3}m}$

Cell parameter: $\boldsymbol{a}$ = 7.7255 Å, **Z** = 4

| Atom | Wyckoff site | *x* | *y* | *z* |
|---|---|---|---|---|
| Pr | *8c* | 0.25 | 0.25 | 0.25 |
| Zn | *4a* | 0 | 0 | 0 |
| Mn | *4b* | 0.5 | 0.5 | 0.5 |
| O | *24e* | 0.19450 | 0 | 0 |

↓

$P2_1/n$

$$\left(\begin{array}{ccc|c} ½ & ½ & 0 & ¼ \\ -½ & ½ & 0 & ¼ \\ 0 & 0 & 1 & 0 \end{array}\right)$$

↓

Low symmetric structure

Monoclinic $\boldsymbol{P2_1/n}$

Cell parameters: $\boldsymbol{a}$ = 5.4578Å, $\boldsymbol{b}$ = 5.5295Å, $\boldsymbol{c}$ = 7.7255Å, $\boldsymbol{\beta}$ = 90.089°, and **Z** = 2

| Atom | Wyckoff site | *x* | *y* | *z* | BVS |
|---|---|---|---|---|---|
| Pr | *4e* | 0.50608 | 0.54440 | 0.24672 | 2.884 |
| Zn | *2c* | 0 | 0.5 | 0 | 2.153 |
| Mn | *2d* | 0.5 | 0 | 0 | 3.883 |
| O1 | *4e* | 0.19450 | 0.19931 | -0.04135 | 1.957 |
| O2 | *4e* | 0.23037 | 0.75531 | -0.01892 | 1.979 |
| O3 | *4e* | 0.41526 | 0.02947 | 0.26037 | 1.966 |

| Zn–O1 = 2.00(9) | Mn–O1 = 2.02(10) | Zn–O1–Mn = 150(4) |
|---|---|---|
| Zn–O2 = 1.91(11) | Mn–O2 = 1.99(11) | Zn–O2–Mn = 169(5) |
| Zn–O3 = 1.92(6) | Mn–O3 = 2.07(6) | Zn–O3–Mn = 151(3) |

$R_{exp}$ = 6.95, $R_p$ = 5.46, $R_{wp}$ = 6.97 and $\chi^2$ = 1.01

**TABLE 2**

| Band no. | Frequency ($cm^{-1}$) | FWHM ($cm^{-1}$) | Symmetry |
|---|---|---|---|
| 1 | 94 | 21 | $T$ ($B_g$) |
| 2 | 107 | 2 | $T$ ($B_g$) |
| 3 | 117 | 18 | $T$ ($B_g$) |
| 4 | 131 | 24 | $T$ ($A_g$) |
| 5 | 140 | 3 | $T$ ($A_g$) |
| 6 | 154 | 24 | $T$ ($A_g$) |
| 7 | 168 | 30 | $L$ ($B_g$) |
| 8 | 193 | 44 | $L$ ($B_g$) |
| 9 | 235 | 67 | $L$ ($B_g$) |
| 10 | 281 | 32 | $L$ ($A_g$) |
| 11 | 320 | 93 | $L$ ($A_g$) |
| 12 | 388 | 39 | $v_5$ ($B_g$) or $L$ ($A_g$) |
| 13 | 429 | 46 | $v_5$ ($B_g$) |
| 14 | 479 | 168 | $v_5$ ($A_g$) |
| 15 | 541 | 65 | $v_2$ ($B_g$) |
| 16 | 608 | 101 | $v_2$ ($B_g$) |
| 17 | 672 | 35 | $v_2$ ($A_g$) |
| 18 | 797 | 139 | $v_1$ ($B_g$) |
| 19 | 858 | 72 | $v_1$ ($A_g$) |

## Table Captions

TABLE 1: Refined structural parameters of PZM.

TABLE 2: Observed Raman active phonon modes of PZM.